\documentclass[12pt,letterpaper]{article}
\usepackage[utf8]{inputenc}
\usepackage[english]{babel}
\usepackage{amsmath}
\usepackage{amsfonts}
\usepackage{amssymb}
\usepackage{graphicx} 
\usepackage{listings}
\usepackage{color}
\usepackage{float}
\usepackage{textgreek}
\usepackage[left=2cm,right=2cm,top=2cm,bottom=2cm]{geometry}
\usepackage[tagged]{accessibility}
\usepackage[pdflang={en-US}]{hyperref}
\usepackage{xcolor}
\usepackage{siunitx}
\DeclareSIUnit\year{yr}

\allowdisplaybreaks

\newcommand{\authorstyle}[1]{{\large\usefont{OT1}{phv}{b}{n}\color{black}#1}} 

\newcommand{\institution}[1]{{\footnotesize\usefont{OT1}{phv}{m}{sl}\color{black}#1}} 

\usepackage{titling} 

\newcommand{\HorRule}{\color{gray}\rule{\linewidth}{1pt}} 

\pretitle{
	\vspace{-30pt} 
	\HorRule\vspace{10pt} 
	\fontsize{24}{36}\usefont{OT1}{phv}{b}{n}\selectfont 
	\color{black} 
}

\posttitle{\par\vskip 15pt} 

\preauthor{} 

\postauthor{ 
	\vspace{-10pt} 
	\par\HorRule 
	\vspace{20pt} 
}

\title{Epidemiology of the Living Dead: A social Force Model of a Zombie Outbreak}

\author{
	\authorstyle{Sydney Balkovitz, Alyssa Croco, Jake Garda, Maggie Hatch, Franklyn Paul, Lauren Vu, Tristan West and Gavin Buxton} 
	\newline\newline 
	\institution{Science Department, Robert Morris University, Moon Township, PA 15108, US.}\\ 
}

\date{}

\begin{document}

\maketitle 

\vspace{-80pt}

We adapt the social force model of crowd dynamics to capture the evacuation during a zombie outbreak from an academic building. 
Individuals navigate the building, opening doors, and evacuate to the nearest exit.
Zombies chase the uninfected individuals, and once caught there is a probability of a susceptible individual being infected or killed, or for the zombie to be killed by the person being attacked.
We find that the speed of the zombies plays a crucial role in the dynamics of the evacuation, the rate of infection, and the number of casualties during the outbreak.
The model leads to insights that may be relevant to other, less fictitious, emergency situations.

\section*{Introduction}

As Zombies have gained traction in popular culture, they have also found an important application in public health. 
Models of zombie apocalypses offer a lens through which researchers can examine the potential danger of disease outbreaks. 
For example, a zombie infestation has often been correlated with a rabies endemic, enabling epidemiologists to spread awareness about the disease in an engaging manner \cite{nasiruddin2013zombies}. 
In a disaster forum on twitter, the CDC found that some of the public compared their readiness for a natural disaster to that of a zombie apocalypse \cite{nasiruddin2013zombies}. 
Although fictitious, the concepts of zombie infection transmission, the psychological response, and potential mitigating outbreak strategies offer insights and guidelines for addressing real-world public health concerns. 
Insights into the dynamics of disease transmission, the importance of different medical responses, and the effects of policies and mitigating strategies can be conveyed by considering the similarities between real disease outbreaks and the epidemiology of a zombie outbreak \cite{munz2009zombies, guitton2014does, alemi2015you}.
Community outreach in both public health and science pedagogy have begun to use zombie pop culture to enhance the education of critical health issues.

The modern day zombie can be traced back to voodoo practitioners and the inhumane conditions of the Haiti slaves' subjugation.
Zombie powders developed by voodoo practitioners contained tetrodotoxin, a deadly neurotoxin, that resulted in difficulty walking and mental confusion; in higher doses temporary paralysis may occur, with victims being buried alive and later arising from apparent death \cite{mariani2015tragic, hines2008zombies}.
George A. Romero inadvertently invented modern day zombies in the Night of the Living Dead. 
These zombies were characteristically slow, but in later films (such as 28 Days Later and World War Z) zombies started to run and relentlessly pursue their victims and even climb walls.
The zombie behavior that are displayed in different movies (and other media) are, therefore, quite varied.

In terms of real zombies, there has been a wide range of diseases that may exhibit zombie-like behavior.
Prion diseases (also known as transmissible spongiform encephalopathy) are transmissible, untreatable, fatal brain diseases in mammals \cite{johnson2005prion}. Symptoms included rapidly developing dementia, difficulty walking and changes in gait, sudden uncontrollable movements of muscles, hallucinations and confusion; symptoms that may be described as zombie-like.
Parasites and parasitoids may also manipulate the behavior and physiology of an infected host. 
For example, fungi within the genus Ophiocordyceps can infect ants, hijacking the central nervous system, and manipulating its victims to exhibit zombie-like behavior. 
In particular, the ant is manipulated to climb to a location that will facilitate fungal spore dispersal, before the fungus kills and grows inside the host, and scatters its spores from the rupturing head of its host \cite{tang2023six, lin2020evaluating}.
Rodents infected with toxoplasma gondii, develop parasitic cysts within their neurons and glia, that makes them attracted to the odor of cat urine and display erratic behaviors that will lead them to being caught by a cat, increasing the parasite transmission to feline hosts \cite{gale2021examining}.
That said, there are no diseases with a short incubation period that spreads by forcing an infected individual to exhibit homicidal tendencies towards susceptible individuals; any potential zombie apocalypses, therefore, remains fictitious. 

Here we adopt the social force model for capturing the movement of individuals in an academic building consisting of several classrooms, a dozen science laboratories, over 20 faculty offices, four bathrooms and various storage or laboratory preparation rooms. The social force model is a particular example of agent based modeling that captures the movement of individuals in response to physical and social forces \cite{helbing1995social, helbing2000simulating}. In particular, the model may account for the social desire of individuals to give one another personal space and the physical forces as crowding occurs. These forces result in the movement of individuals through the integration of Newton's second law of motion. Such models have been used to capture crowding \cite{helbing2000simulating}, and evacuations in various emergencies including fire \cite{cao2023development}, active shooting situations \cite{anklam2014mitigating, arteaga2020building} and shipwrecks \cite{kang2019improved}. 
Most social force models of evacuations have considered relatively simple building configurations (simple rectangular rooms); although more complicated configurations have been considered \cite{ha2012agent, anklam2014mitigating, arteaga2020building, ma2021optimization, alac2023optimising}.  The implementation of more complex building configurations have enabled such models to optimize the location of exits in school buildings \cite{ma2021optimization, alac2023optimising},  and elucidate the effects of building design on egress times \cite{arteaga2020building}. 
Coupling the social force model to epidemiology models has recently been used to elucidate the effects of crowd evacuation dynamics on the spread of an infectious disease \cite{agnelli2023spatial}.
In particular, they found that the spread of respiratory infectious diseases can be exacerbated by the crowding and gathering formations within indoor venues, and discussed possible policies to increase social distancing and reduce contagion.
Fear and anxiety has also been included in social force models using epidemiology models, where individuals can become infected with fear and panic \cite{cornes2019fear}. In the current model, we capture the spread of a zombie infection from zombies to susceptible individuals, and their transition from infected to zombies, as individuals attempt to evacuate from the building where the initial outbreak occurs.

\section*{Model}

The agent based modeling employed in the current study adopts the social force model of Hellbing \emph{et al.} \cite{helbing1995social, helbing2000simulating, helbing2005self, moussaid2011simple}. In other words, the evacuating individuals (and zombies) are modeled as point masses subject to forces that mimic social and physical interactions with an individual and their surroundings. In particular, the following force acts on a person
\begin{equation*}
\vec{f}_{\alpha} = \frac{m_{\alpha}}{\tau_{\alpha}} \left( v_{\alpha}^0 \vec{e}_{\alpha} - \vec{v}_{\alpha}\right) + \sum_{\beta \neq \alpha} \vec{f}_{\alpha\beta} + \sum_{\alpha w} \vec{f}_{\alpha w} + \sum_{\alpha z} \vec{f}_{\alpha z}
\end{equation*}
Where the first term captures a persons movement towards a desired velocity. The person has a desired speed, $v_{\alpha}^0$, and a desired direction, $\vec{e}_{\alpha}$. The current velocity of an individual is given by $\vec{v}_{\alpha}$, their mass is, $m_{\alpha}$, and the parameter $\tau_{\alpha}$ is the timescale over which the individual might change their velocity to the desired one. The second term captures the physical interaction between two individuals (individuals represented by subscripts $\alpha$ and $\beta$). The third term captures the interaction between an individual and the walls, and the final term accounts for the interactions between zombies and other individuals.

The geometry of the building is depicted in Fig. 1. The walls are represented by gray lines, the exits are represented by red lines and internal doors (which may be open or closed) are represented by blue lines. The desired direction that the non-zombie individuals will attempt to move in will be the direction towards the nearest exit. Waypoints are distributed throughout the buildings geometry and waypoints are considered connected if they are in line of sight of one another (the line between the waypoints does not cross a wall). The shortest distance to each exit is established via the network of connected waypoints, and an individual will move towards the waypoint whose distance (distance to waypoint and distance from waypoint to the exit) is shortest \cite{lozano1979algorithm}. As an example, Fig. 1 depicts the waypoints as circles whose color represents the distance from the waypoint to the main entrance. Waypoints are typically placed at crossroads in corridors and inside/outside of doorways to facilitate a network of points that optimize the egress from the building \cite{parisi2023we}.

The desired speed of individuals increases when an individual is panicked. In particular, the individual becomes panicked when in proximity to a zombie and the degree of panic (and increased speed) exponentially reduces with time. The zombies in the model are not attracted to exits, but to the closest noninfected individual. The noninfected individuals have to be in line of sight of the zombie in order for the zombies' desired velocity to be in the direction of the  noninfected individual.

$\vec{f}_{\alpha\beta}$ is the force between two individuals and is given by
\begin{equation*}
\vec{f}_{\alpha\beta} = k_1 g(2 R - d_{\alpha\beta}) \vec{n}_{\alpha\beta} + k_2  g(2 R - d_{\alpha\beta}) \Delta v_{\beta \alpha}^t \vec{t}_{\alpha\beta}
\end{equation*}
where this function does not include soft repulsive interactions, as during the evacuation such social niceties, and the respect for personal space, are assumed to play a less significant role than the desire to evacuate quickly. The constants $k_1$ and $k_2$ control the interactions between individuals as they overlap and slide past one another, respectively.
The function $g(x)$ is equal to the argument $x$ when individuals are in contact ($2 R - d_{\alpha\beta} > 0$), and is otherwise zero. $R$ is the radius of an individual and $d_{\alpha\beta}$ is the distance between two individuals (individuals represented by subscripts $\alpha$ and $\beta$). The normal vector pointing from one individual to another is given by
\begin{equation*}
\vec{n}_{\alpha\beta} = (\vec{r}_{\alpha} - \vec{r}_{\beta})/d_{\alpha\beta}
\end{equation*}
and a vector is defined in the tangential direction through the \qty{90}{\degree} rotation of the normal vector.
\begin{equation*}
\vec{t}_{\alpha\beta} = R_{\qty{90}{\degree}} \vec{n}_{\alpha\beta}
\end{equation*}
The relative velocity of two individuals as they slide past one another is then given by
\begin{equation*}
\Delta v_{\beta \alpha}^t = (\vec{v}_{\beta} - \vec{v}_{\alpha}) \cdot \vec{t}_{\alpha \beta}
\end{equation*}
The above functional form inhibits people from overlapping in space and from sliding past one another when they are in close confinement (e.g., at a doorway or other bottleneck). A similar functional form is adopted for the wall
\begin{equation*}
\vec{f}_{\alpha w} = \left\{ A_w \exp \left[ \frac{R - d_{\alpha w}}{B_w}\right] + k_1 (R - d_{\alpha\beta})\right\} \vec{n}_{\alpha w} + k_2 g(R - d_{\alpha\beta}) (\vec{v}_{\alpha} \cdot \vec{t}_{\alpha w}) \vec{t}_{\alpha w}
\end{equation*}
where $A_w$ is the strength, $B_w$ the range, of a soft repulsive force. $d_{\alpha w}$ is the distance from an individual to the nearest point of the wall. The vectors $\vec{n}_{\alpha w}$ and $\vec{t}_{\alpha w}$ are the normal and tangential directions. The normal vector points towards the nearest point on the wall and the tangential is obtained by rotating the normal vector by \qty{90}{\degree}.  

The repulsive force acting on individuals in close proximity to a zombie is of the form
\begin{equation*}
\vec{f}_{\alpha z} = k_z g(R_z - d_{\alpha z}) \vec{n}_{\alpha z}
\end{equation*}
where $k_z$ is the strength of this repulsion, $R_z$ is the range of this interaction, and $d_{\alpha z}$ is the distance from an individual (not zombie) and a zombie. The vector $\vec{n}_{\alpha z}$ is a unit vector pointing from the zombie to the individual.
Once the social and physical forces acting on an individual are calculated, Newton's second law is integrated to obtain their updated velocities and positions.
\begin{equation*}
m_{\alpha} \frac{d \vec{v}_{\alpha}}{d t} = \vec{f}_{\alpha}
\end{equation*}
In this manner the individuals are able to navigate and interact with their surroundings, and each other, as they head towards their objectives; for non-zombies that is to exit the building in the current model, and for zombies that is to attack the nearest non-infected individual that is in their line of sight. 

The individuals can interact with the surroundings in the current model mainly through opening and closing doors. Individuals have a randomly assigned capability to open doors that are initially locked with a probability, $P_{open}$ (faculty, staff, and student workers, for example). Individuals that spawn in locked rooms are also assumed to fall in this category. Therefore, when calculating the closest waypoint for an individual to travel towards when exiting the building, whether or not the individual can open doors, and whether someone else has opened a door, has to be taken into consideration. The distances are recalculated every \qty{2}{\s} in the current model as this is assumed to be a reasonable time frame for an individual to change their mind, and decide to exit in a more opportune direction. In other words, not all individuals are exiting towards the same exits, even when spawned in similar locations, as individuals without the ability to open locked doors may not be able to access the closest exit.

If an individual is within a given distance, $d_z$, of a zombie then their is a probability, $P_{catch}$ that the zombie will grab hold of the individual. If the zombie catches the individual then there is a probability, $P_{infect}$, that the zombie will turn the individual from susceptible to infected, a probability, $P_{kill}$, that the zombie will kill the individual, and a small probability $P_{killz}$ that the individual will kill the zombie. If an individual is infected then the zombie will no longer attack them and they will have an average incubation period of $t_{inc}$, after which time the individual will become zombified.

The building shown in Fig. 1 is a unique configuration for a science and engineering building. The building was once a student recreational center, and still has a large gym to the left of the building. 
To the left of the gym are the main classrooms. 
Most other large rooms in the building are science and engineering laboratories. 
The initial distribution of people, and which doors are open/closed at the beginning of a simulation, is dictated by the day of the week. 
Lecture classes are typically Monday, Wednesday and Friday (configuration 1) and laboratory courses are more likely to be offered Tuesday and Thursday (configuration 2). Therefore, the classrooms are more occupied in configuration 1, and laboratories are more occupied (and unlocked) in configuration 2. The question is how, in the event of an evacuation during the early stages of a zombie apocalypse, do people exit the building and how will the type of zombie affect the number of casualties, the infection rates and evacuation times. 

\section*{Results}

The simulations start with 150 people and one zombie. The parameters adopted for the current simulations are given in Table 1. The velocity of the individuals is considered to double when an individual is panicked (within \qty{3}{\m} of a zombie). The time, $\tau_{\alpha}$, over which an individual is considered capable of changing direction is taken to be relatively small in the current simulations as this facilitates individuals traversing doorways. The incubation period is short in the current model, as the simulated time over which people exit the building is relatively short; this might correspond to more modern representations of zombie infections (e.g., 28 Days Later). 

Figure 2 depicts the interaction of a zombie and a classroom of susceptible individuals in the current model. The zombie is colored green, while uninfected individuals are black. Fig. 2a. shows the initial random distribution of people at the beginning of the simulation.  The situation 1 s later is shown in Fig. 2 b. The zombie is chasing two individuals into the bottom left corner of the room. Fig. 2c. shows that one of the individuals is now infected (colored gray) and the zombie is chasing the second individual towards the door, which is congested with other individuals trying to exit the classroom. After only 3 s the zombie has engaged the individuals by the doorway and infected two other people (the gray circles near the zombie). In Fig. 2e (after only 4 s) the zombie has killed two other people, and has itself been killed. The dead people in the simulation are colored red. After 5 s (Fig. 2f) the initially infected individual has turned into a zombie and has begun chasing susceptible individuals, and most people have left the room. In this example the zombies were considered to be fast (with a desired velocity of 4 m/s) and the situation evolves quite quickly, with susceptible individuals becoming panicked when in close proximity to the zombie, and exiting the room relatively quickly. The fact the system evolves so quickly may be similar to other emergencies, such as active shooter situations or knife attacks.

The times of egress from the building are plotted in Figure 3. Fig. 3a depicts the egress times for configuration 1 (when faculty and students are mostly in classrooms) and Fig. 3b depicts egress times for configuration 2 (when faculty and students are mostly in laboratories). The cumulative distribution function (integration of probability distribution function) of the egress times for systems with different zombie velocities are contrasted. Most people will have exited the building between 30 and 60 s in the current model; although in reality there may be a delay in peoples reactions and the egress times may be larger. However, there is a noticeable decrease in egress times in systems with faster zombies. For example, the average time for people to exit the building when the zombies are faster is \qty{8.9}{\s} while for slow moving zombies the average egress time is \qty{14.9}{\s}. As more people are brought into contact with the zombies, and more people become panicked, they exit the building at a faster rate. This disparity is more prominent when the zombie outbreak occurs in a small space (i.e., a classroom) and more susceptible individuals become panicked; despite congestion in smaller spaces being more likely to increase egress times. 

The average numbers of zombies, infected individuals and dead people as a function of time are plotted in Figure 4 for configuration 1. The numbers are averaged over 50 simulations and the standard deviations are not shown as they are relatively large (comparable to the averages). Fig. 4a depicts the systems with slow zombies  (desired speed of \qty{1}{\m\per\s}) and exhibits a gradual increase in the number of zombies and dead individuals.
Fig. 4b depicts the systems with fast zombies (desired speed of \qty{4}{\m\per\s}) and exhibits faster initial growth in the number of infected individuals, zombies and dead people, before saturating at around \qty{30}{\s}. As expected, the number of infected individuals shows an early increase, especially for systems with faster zombies.  The number of deceased individuals at the end of the simulation is also generally greater in systems where the zombies are faster. The ability for individuals to fight back keeps the number of zombies at a limited number, and in the current model the number of zombies does not appear to grow uncontrollably. 

Similarly, Figure 5 depicts the average numbers of zombies, infected individuals and dead people as a function of time for configuration 2 (when students are initially more distributed in laboratories and less so in classrooms). However, in the less crowded confines of the laboratories a zombie outbreak is less likely to spread, especially when the zombies are slower moving. Fig. 5a depicts the systems with slower moving zombies and the average number of infected individuals remains less than one, and the average number of dead people (potentially including the zombie) is around one. Contrast this with the first configuration (where people are more concentrated in classrooms) where the number of dead people reached more than 3 on average in systems that also consisted of slower zombies. Fig. 5b depicts systems with faster zombies (desired speed of \qty{4}{\m\per\s}) and the behavior is similar to Fig. 4b., but with smaller numbers of infected individuals, zombies and dead people. Again, this is a consequence of the zombie outbreak not occurring in more populated classrooms and occurring in laboratories.

\section*{Conclusions}

A social force model has been developed to capture the evacuation from a large academic building. The model is used to capture the initial outbreak during a potential zombie apocalypse. In the current model, the zombies are particularly adept at cornering susceptible individuals (even when slow). It is found that the faster the zombies are capable of moving (more similar to 28 Days Later or World War Z, and less like Night of the Living Dead zombies) the shorter the egress times (as panicked individuals travel faster), the faster the spread of infection, and the more casualties that are expected to occur. In addition, it was found that the outcome of the zombie outbreak might be more severe on a Monday, Wednesday, or Friday when faculty and students are more concentrated in classrooms than on Tuesdays or Thursdays when faculty and students are spread between various laboratories. 

In the current model we kept the parameters for the individuals (e.g., mass and desired speed) independent of the individual. Most studies might assign a Gaussian distribution to such parameters; although in studies with a fixed population (such as college students) the parameters might not be as variable. 
Another aspect of the model that is missing is the social bonds that exist between friends. Friends may seek comfort and be more likely to exit alongside their friends \cite{helbing1995social}. Such attractive interactions are considered to be less influential in a university setting, than in other social gatherings with family members.
However, given that students (especially in the US) are increasingly trained for active shooting situations, some of that training may come into play during a zombie apocalypse. Future work could include the standard run, hide and fight advise given to students; where here the individuals are limited to running and fighting (as a last resort) in their responses.

In conclusion, we show that the social force model provides an efficient method for predicting the response of individuals to an emergency in a complex environment. For the first time, we have applied the model to evacuation dynamics in a university building, included the ability of individuals to interact with their surroundings and open doors, and utilized this model to predict the outcome of a zombie outbreak. The application of this model to a zombie outbreak will hopefully provide a framework for understanding and modeling other emergency situations.

\clearpage
\section*{Figures and Tables}

\begin{center}
\includegraphics[width=\linewidth]{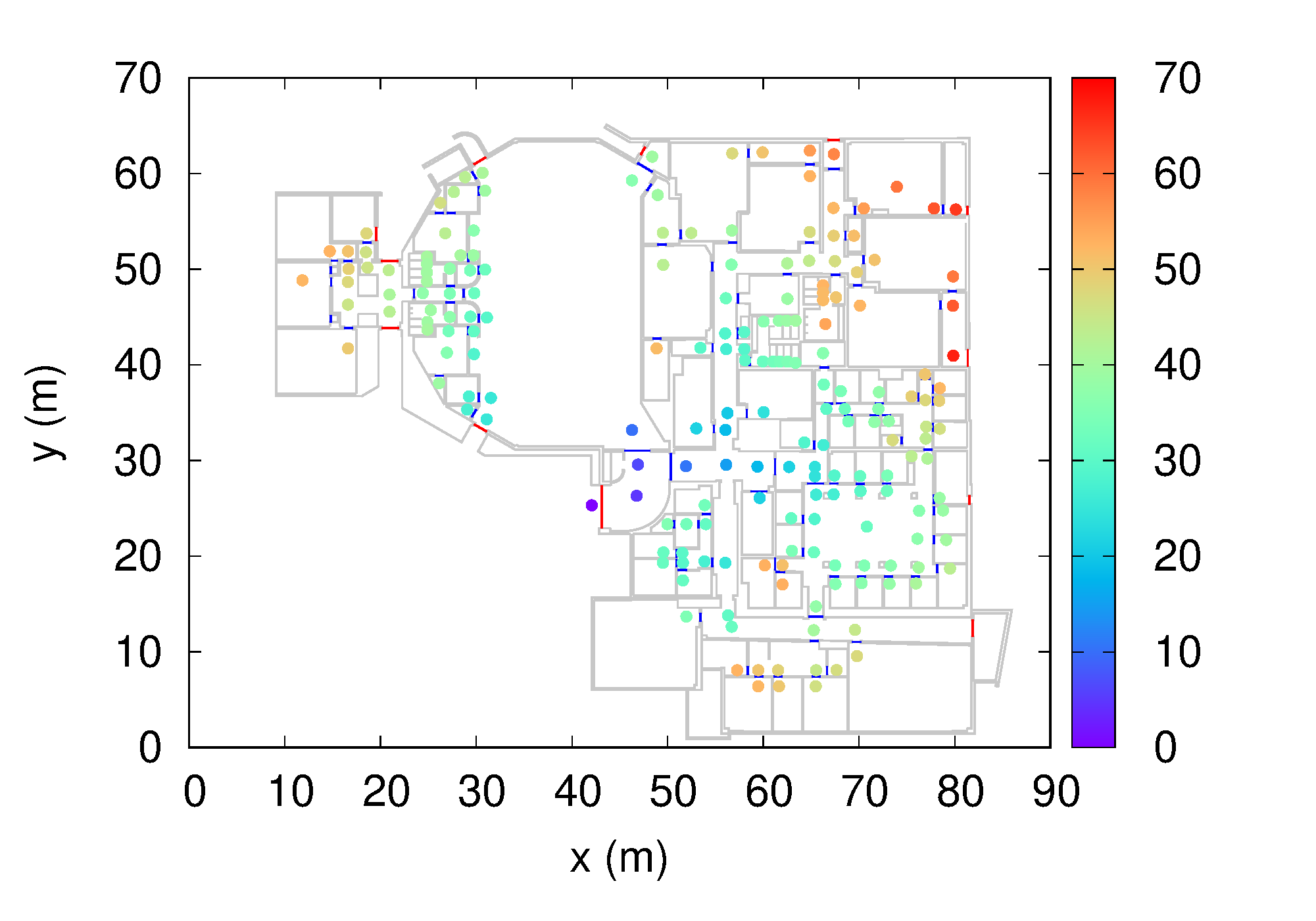}
\end{center}
Figure 1. The geometry adopted in the current model. Gray lines represent walls, red lines the exits and blue lines the interior doorways. The colored circles represent the waypoints in the model and are colored based on the distance to the main entrance. 

\clearpage

\begin{table}[H]
	\begin{center}
	\begin{tabular}{cc} \hline 
	\thead{Parameter} & \thead{Value} \\ \hline 
	$v_{\alpha}^0$ & \qty{1.5}{\m\per\s} \\ 
	$m_{\alpha}$ & \qty{80}{\kg} \\ 
	$\tau_{\alpha}$ & \qty{0.1}{\s} \\ 
	$k_1$ & \qty{1.2e5}{\N\per\m} \\ 
	$k_2$ & \qty{2.4e4}{\N\per\square\m\per\s} \\ 
	$R$ & \qty{0.2}{\m} \\ 
	$A_w$ & \qty{400}{\N} \\ 
	$B_w$ & \qty{0.1}{\m} \\ 
	$k_z$ & \qty{1e4}{\N} \\ 
	$R_z$ & \qty{2.9}{\m} \\ 
	$P_{open}$ & 0.2 \\ 
	$d_z$ & \qty{0.6}{\m} \\ 
	$P_{catch}$ & 0.8 \\ 
	$P_{infect}$ & 0.8 \\ 
	$P_{kill}$ & 0.1 \\ 
	$P_{killz}$ & 0.15 \\ 
	$t_{inc}$ & \qty{10}{\s} \\ \hline 
	\end{tabular} 
	\end{center}
 \end{table}

Table 1. The values used for parameters in the current simulations.

\clearpage

\begin{center}
\includegraphics[width=\linewidth]{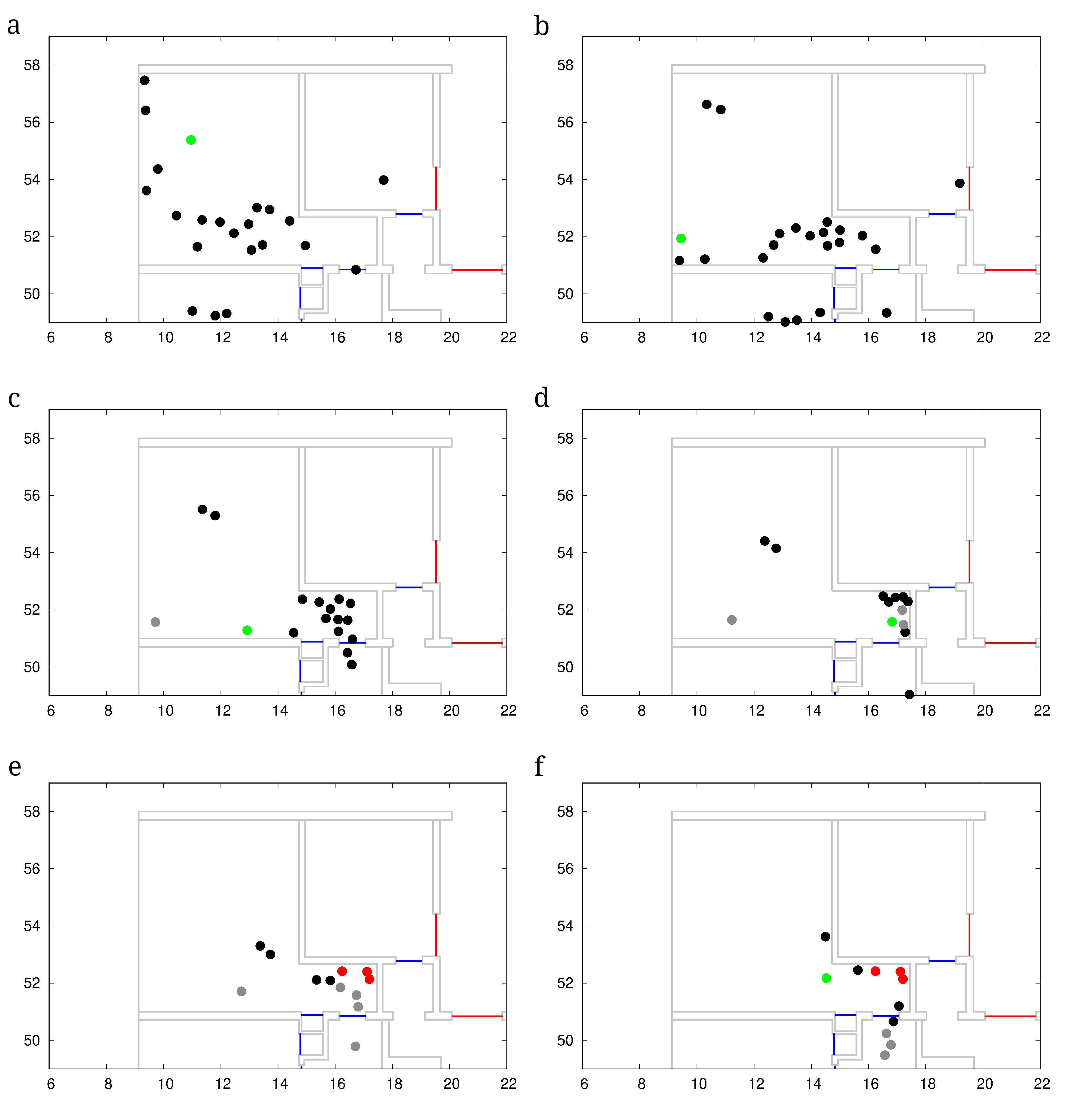}
\end{center}
Figure 2. An example of the simulation of a zombie outbreak in a classroom. Susceptible individuals (black circles), infected individuals (gray circles), zombies (green circles) and dead people (red circles) are shown after a) \qty{0}{\s}, b) \qty{1}{\s}, c) \qty{2}{\s}, d) \qty{3}{\s}, e) \qty{4}{\s}, and f) \qty{5}{\s}.  The zombie chases after and infects susceptible individuals, kills susceptible individuals and is itself killed, before an infected individual turns into a zombie. 

\clearpage

\begin{center}
\includegraphics[width=0.8\linewidth]{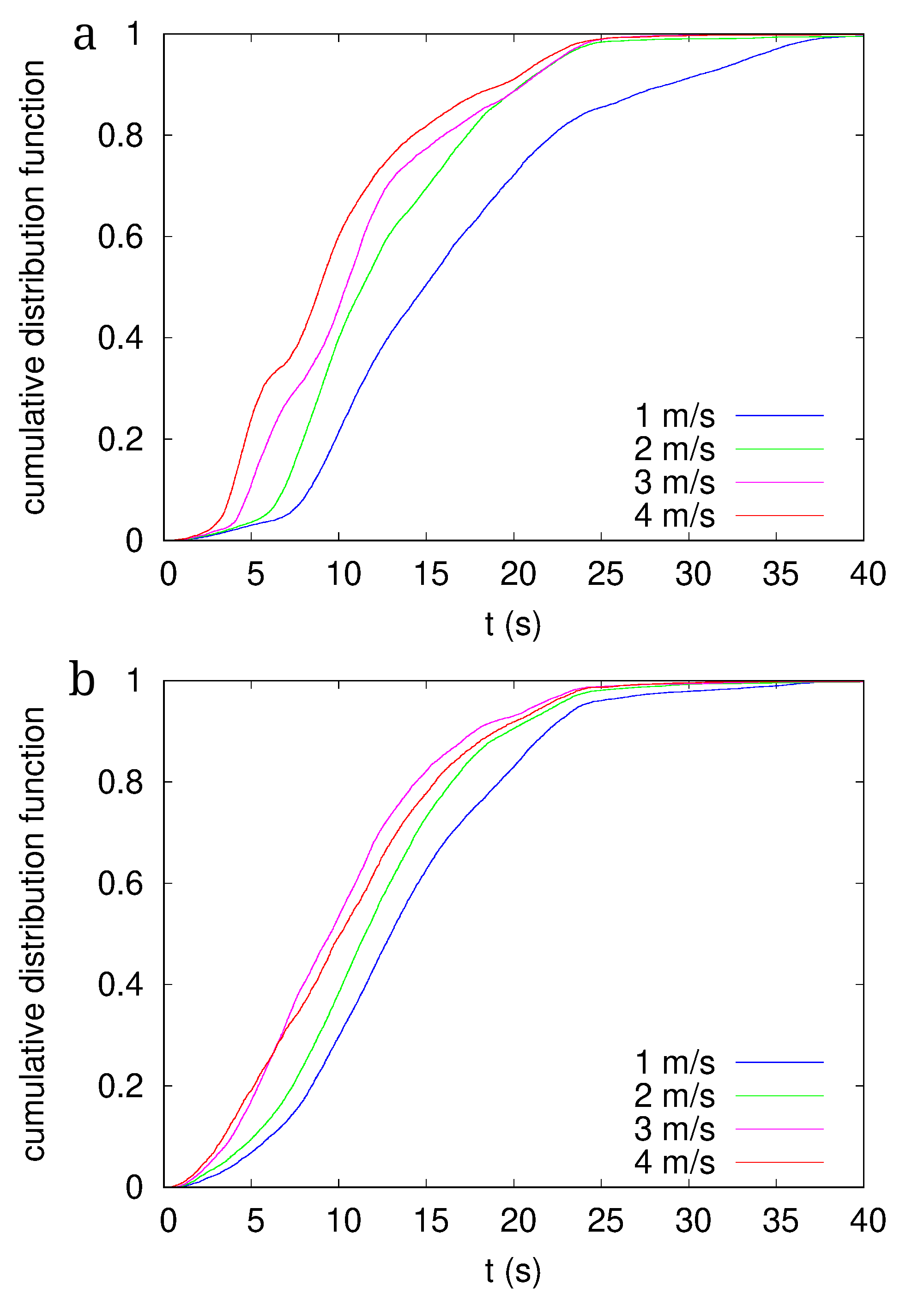}
\end{center}
Figure 3. The cumulative distribution function of egress times for a) configuration 1 (where most faculty and staff are in classrooms) and b) configuration 2 (where most faculty and staff are in laboratories). The effects of increasing the zombie velocity is shown to reduce egress times.

\clearpage

\begin{center}
\includegraphics[width=0.8\linewidth]{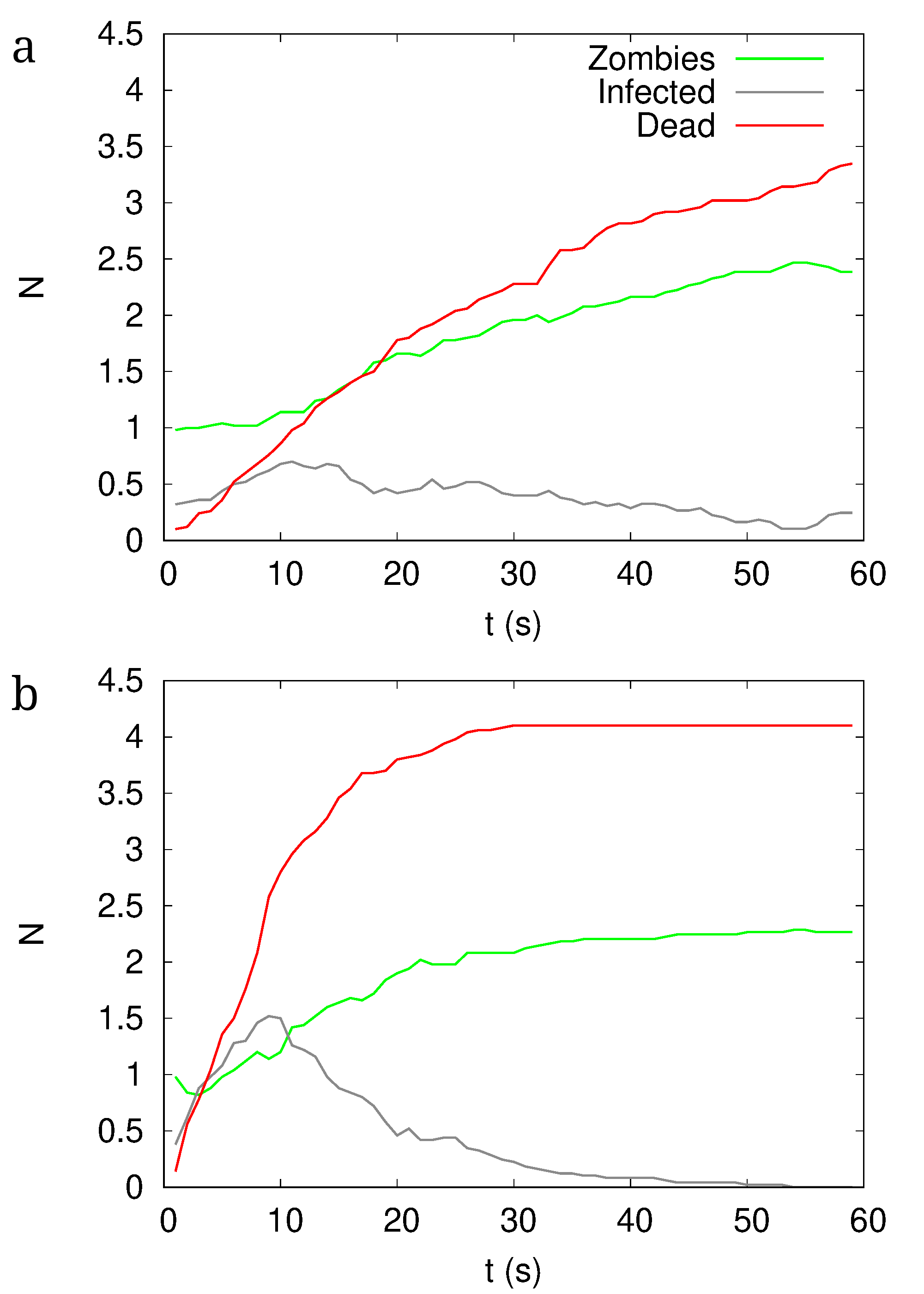}
\end{center}
Figure 4. The number of zombies, infected individuals and dead people as a function of time. Data is averaged over 50 simulations. Systems consisting of a) slow moving zombies with a speed of \qty{1}{\m\per\s} are contrasted with systems consisting of b) fast moving zombies with a speed of \qty{4}{\m\per\s}. The starting configuration corresponds with most faculty and students in classrooms (configuration 1). 

\clearpage

\begin{center}
\includegraphics[width=0.8\linewidth]{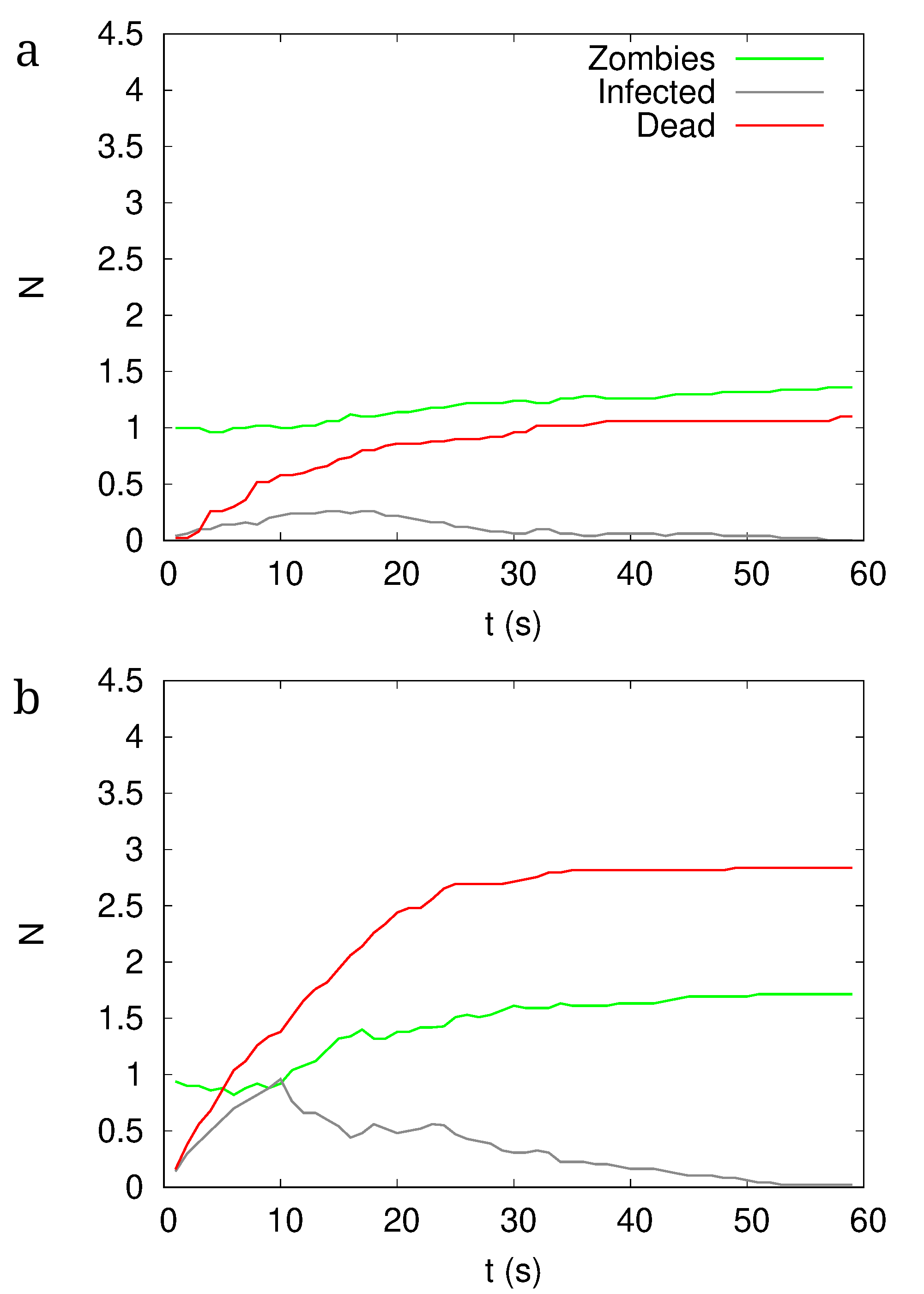}
\end{center}
Figure 5. The number of zombies, infected individuals and dead people as a function of time. Data is averaged over 50 simulations. Systems consisting of a) slow moving zombies with a speed of \qty{1}{\m\per\s} are contrasted with systems consisting of b) fast moving zombies with a speed of \qty{4}{\m\per\s}. The starting configuration corresponds with most faculty and students in laboratories (configuration 2).

\clearpage
\bibliographystyle{unsrt} 
\bibliography{refs.bib}

\end{document}